\documentclass[prb,aps,showpacs,floatfix,floats,nobalancelastpage,twocolumn]{revtex4}
\usepackage{graphicx}
\usepackage{amsmath}
\usepackage{color}
\usepackage{dcolumn}
\usepackage{bm}

\def\etal{~\textit{et~al}}
\def\ra{\rangle}
\def\la{\langle}

\def\hc{{\rm h.c.}}
\def\CsCuBr{{Cs$_2$CuBr$_4$}}

\begin{document}

\title{Variational Studies of Triangular Heisenberg Antiferromagnet in Magnetic Field}
\author{Tiamhock Tay}
\affiliation{Department of Physics, California Institute of Technology, Pasadena, CA 91125}

\author{Olexei I. Motrunich}
\affiliation{Department of Physics, California Institute of Technology, Pasadena, CA 91125}
\date{\today}
\pacs{}


\begin{abstract}
We present a variational study of the Heisenberg antiferromagnet on the spatially anisotropic triangular lattice in magnetic field.  First we construct a simple yet accurate wavefunction for the 1/3-magnetization plateau $uud$ phase on the isotropic lattice. Beginning with this state, we obtain natural extensions to nearby commensurate coplanar phases on either side of the plateau.  The latter occur also for low lattice anisotropy, while the $uud$ state extends to much larger anisotropy.  Far away from the 1/3 plateau and for significant anisotropy, incommensurate states have better energetics, and we address competition between coplanar and non-coplanar states in the high field regime.  For very strong anisotropy, our study is dominated by quasi-1d physics.  The variational study is supplemented by exact diagonalization calculations which provide a reference for testing the energetics of our trial wavefunctions as well as helping to identify candidate phases.
\end{abstract}
\maketitle

\section{Introduction}

The spin-1/2 Heisenberg antiferromagnet on a spatially anisotropic 2d triangular lattice is a deceptively simple spin system which nevertheless possesses very rich physics.  Despite having attracted much attention, a complete understanding of the model has not been achieved.  This can be attributed to the enhanced quantum fluctuations arising from a combination of low dimensionality, small spin, geometrical frustration and spatial anisotropy, thus leading to a rich phase diagram.  At zero field, studies have suggested that the anisotropic system may possibly remain disordered even at zero temperature.  In experimental realizations of the triangular antiferromagnet, a 1/3-magnetization plateau was found for the approximately isotropic material $\textrm{Cs}_2\textrm{CuBr}_4$ but not for the more anisotropic $\textrm{Cs}_2\textrm{CuCl}_4$.\cite{Ono03,Ono04,Ono05,Fujii04,Fujii07,Tsujii07}  A recent experimental study of $\textrm{Cs}_2\textrm{CuBr}_4$ further revealed a cascade of phases in the fields above the 1/3 plateau, which are still not understood.\cite{Fortune09}

Analytical studies on the model have been done on specific regions of the phase diagram, for instance low anisotropy near the 1/3-magnetization plateau,\cite{Chubukov91,Alicea09} large anisotropy limit,\cite{Starykh07} and high field limit.\cite{Nikuni95, Veillette06}  Several numerical studies using exact diagonalization,\cite{Bernu94,Honecker04,Yoshikawa04,Miyahara06} series expansion,\cite{Honecker99,Pardini08} coupled-cluster method,\cite{Farnell09} density matrix renormalization group,\cite{Weng06,Jiang09} and variational approaches\cite{Huse88,Heidarian09a,Heidarian09b,Yunoki06} have also been used to analyze the model.  Motivated by the experimental and theoretical works, we perform a variational study using simple yet powerful wavefunctions, attempting to cover a large portion of the phase diagram.

\begin{figure}
  \centering
  \includegraphics{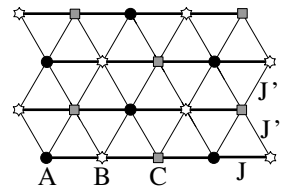}
  \caption{Triangular antiferromagnet with coupling constants $J$ and $J'$ between nearest neighbours along horizontal and oblique directions respectively.  Three sublattices $A$, $B$, and $C$ important in the isotropic case are also labelled.}
  \label{fig:ani_lattice}
\end{figure}

\begin{figure}[t] 
  \centering
  \includegraphics[width=\columnwidth]{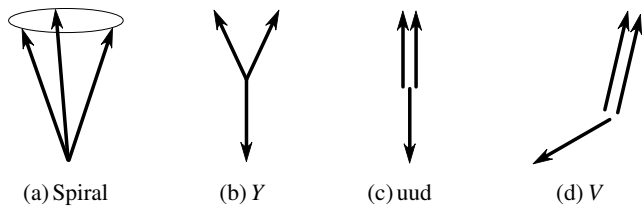}
  \caption{Spin orderings on the isotropic triangular lattice in the field, where three arrows refer to three sublattices indicated in Fig.~\ref{fig:ani_lattice}: a) spiral (non-coplanar umbrella); b,c,d) coplanar $Y$, $uud$, and $V$.}
  \label{fig:spin_orderings}
\end{figure}

\begin{figure}[t] 
  \centering
  \includegraphics[width=\columnwidth]{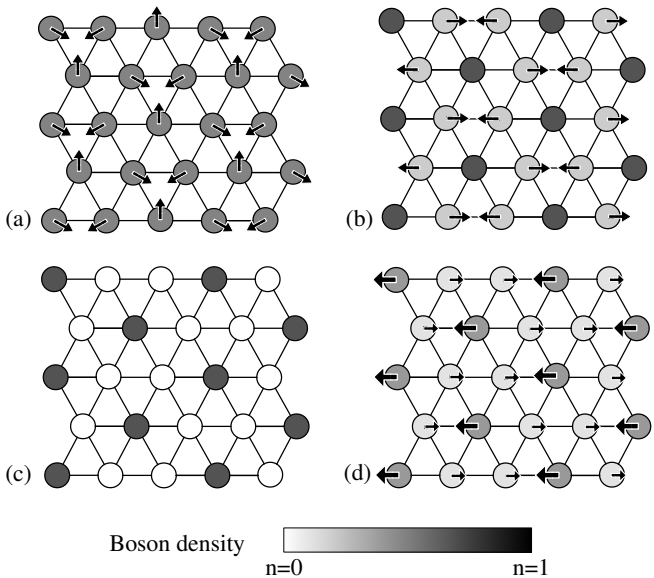}
  \caption{Boson interpretation of the spin orderings in Fig.~\ref{fig:spin_orderings}.  Gray scale shows density order $\la n_r \ra$, while arrows show superfluid order $\la b_r^\dagger \ra$.  a) Spiral is a uniform superfluid phase with rotating phase angles; c) $uud$ is a Mott insulator with $\sqrt{3} \times \sqrt{3}$ CDW order; b,d) $Y$ and $V$ are supersolid phases which contain both charge density and particular superfluid orderings.}
  \label{fig:boson_orderings}
\end{figure}

We consider the Heisenberg model in external magnetic field $h$, 
\begin{eqnarray}
  \hat{H} &=& \sum_{\la rr' \ra} J_{rr'}\ \vec{S}_r \cdot \vec{S}_{r'} 
- h \sum_r S_r^z ~,
\end{eqnarray}
where $\vec{S}_r$ is the spin operator on site $r$ and $J_{rr'}$ are the nearest neighbour exchange couplings.  Throughout, we extensively use hard-core boson picture, mapping $S_r^+ = b_r$ and $S_r^z = \frac{1}{2} - n_r$:
\begin{eqnarray}
  \hat{H} &=& -\sum_{\la rr'\ra} t_{rr'} \left( b_r^\dagger b_{r'} + \hc \right)
+ \sum_{\la rr'\ra} J_{rr'} n_r n_{r'} \label{eqn:Hb}\\
&& - \mu \sum_r n_r + {\rm const} ~, \\
t_{rr'} &=& -\frac{1}{2} J_{rr'}, \quad \mu = -h + \frac{1}{2} \sum_{r' \in r} J_{rr'} ~.
\label{eqn:frustration}
\end{eqnarray}
The boson hopping amplitudes are negative and therefore frustrated on the triangular lattice, making this a challenging interacting problem.

Figure~\ref{fig:spin_orderings} depicts spin ordered phases considered in our variational study.  While these are simple to draw, realizing them as wavefunctions is non-trivial.  The spiral phase is a boson superfluid containing rotating phase angles (see Fig.~\ref{fig:boson_orderings}a) and is captured by an elegant Huse-Elser generalization of the Bijl-Jastrow wavefunction.\cite{Huse88}  For the other phases, constructing simple and yet accurate wavefunctions is not as straightforward and requires consideration of their physical nature in terms of bosons.  Thus, the $uud$ is a Mott insulating phase (see Fig.~\ref{fig:boson_orderings}c) which requires a wavefunction with strongly localized bosons and rapidly decaying correlations.  The $Y$ phase is an interesting supersolid phase (see Fig.~\ref{fig:boson_orderings}b) with rapidly decaying boson correlations between sites on one of the three sublattices as well as long-range correlation between sites on the other two sublattices.  The $V$ phase is a different supersolid (see Fig.~\ref{fig:boson_orderings}d) with long-range boson correlations between all sites; here we find that two different constructions of the trial wavefunctions are required to capture the lower and higher density regimes.  (We remark that supersolid phases of bosons on the half-filled triangular lattice and in the presence of strong repulsion have been of much recent interest, see Refs.~\onlinecite{Wang09,Jiang09,Heidarian09b} and citations therein.)

In section~\ref{sec:isotropic}, we present simple, few-parameter candidate wavefunctions used in our isotropic study.  Encouraged by the accuracy of these candidates, we generalize the wavefunctions to incommensurate versions for our anisotropic study in section~\ref{sec:anisotropic}.  
We conclude with a discussion of the results and implications for $\textrm{Cs}_2\textrm{CuBr}_4$ in section~\ref{sec:conclusion}.

\section{Isotropic Triangular Antiferromagnet: 6$\times$6 Study}\label{sec:isotropic}

In this section, we consider the isotropic triangular Heisenberg antiferromagnet with $J_{rr'}=J$ for all nearest neighbour links.  Beginning at density $n=1/3$, where we have an excellent wavefunction for the $uud$ Mott insulator phase of Fig.~\ref{fig:boson_orderings}c, we construct similarly inspired wavefunctions for nearby supersolid phases of Figs.~\ref{fig:boson_orderings}b,d.  Next, we describe Bijl-Jastrow-type wavefunctions for the spiral of Fig.~\ref{fig:boson_orderings}a and $V$ supersolid of Fig.~\ref{fig:boson_orderings}d, which perform better for $n$ further from 1/3.  We also discuss an alternative construction of the $uud$ state using a $\det \times \det$ (``2-parton'') trial wavefunction.  For the ED calculations, we compute ground state energies for $N_b \leq 12$, while $N_b \geq 13$ data is taken from Bernu\etal.\cite{Bernu94}

We perform all studies at fixed boson number $N_b$.  For each wavefunction below, we also include a Jastrow factor
\begin{eqnarray}
  {\rm Jastrow}(\{n_r\}) = e^{-\frac{1}{2} \sum_{r,r'} u_{rr'} n_r n_{r'}}
\end{eqnarray}
with simple choices of pseudopotentials $u_{rr'}$ providing additional variational freedom.

\subsection{$uud$ state at $n=1/3$}
\label{subsec:uud}

\begin{table}
  \begin{tabular*}{\columnwidth}{@{\extracolsep{\fill}}cccccc}
    \hline
    Trial state & $\mathrm{N_{par}^{orb}}$ & $\mathrm{N_{short}^{Jas}}$ & $\mathrm{N_{long}^{Jas}}$ & $\mathrm{N_{par}^{tot}}$ & E/bond\\
    \hline
    classical &   &   &   &   & -0.0833\\
    spiral  & 0 & 1 & 2 & 4 & -0.1265\\
    $uud$     & 1 & 0 & 0 & 1 & -0.1347\\
              & 1 & 1 & 0 & 2 & -0.1354\\
              & 1 & 2 & 0 & 3 & -0.1355\\
    exact     &   &   &   &   & -0.1361\\
    \hline
  \end{tabular*}
  \caption{Comparison of $uud$ trial energies for different number of short-range Jastrow parameters for 12 bosons on a $6\times 6$ cluster.  Localized orbitals in the permanent extend only to nearest neighbour sites with amplitude $\alpha$, which is single variational parameter in the first listed $uud$ case.  The second $uud$ case has one nearest-neighbor (nnb) Jastrow pseudopotential which is taken to be the same between any pair of nnb sites, while the third case has two such parameters, one for $A-B$ and $A-C$ nnb pairs and the other for $B-C$ nnb pairs, as is appropriate given lattice symmetries of the $uud$ state.  We also show trial energy for the spiral state with 4 variational parameters (same as in table~\ref{table:spiral}); this state performs poorly compared to the $uud$ state.
}
  \label{table:uud}
\end{table}

At density $n=1/3$, the $uud$ phase is stabilized by quantum fluctuations.  We construct a simple boson wavefunction by using $N_b = N/3$ orbitals localized around sites $A_j$, $j = 1 \dots N/3$, from sublattice $A$:
\begin{eqnarray}
|\psi_{uud} \ra &=& \prod_{j=1}^{N/3}
\left(\sum_r \phi_j^{\rm loc}(r) b_r^\dagger \right) |0 \ra ~,
\label{eqn:uud_wf} \\
\phi_j^{\rm loc}(r) &=& \left\{
\begin{array}{ll}
1, & r = A_j\\
-\alpha, & r = \text{neighbour of $A_j$}\\
0, & \text{otherwise}
\end{array}
 \right. ~,
\label{eqn:uud_matrix} \\
\la \left\lbrace r_k \right\rbrace |\psi_{uud} \ra
&=& \mathrm{Perm} \left[ \phi_j^{\rm loc}(r_i) \right] ~.
\label{eqn:uud_perm}
\end{eqnarray}
For $\alpha=0$ this reduces to the classical CDW state with bosons strictly localized on the $A$ sublattice and minimizing the potential energy.  Non-zero $\alpha$ allows bosons to hop to nearest-neighbor sites and gain some kinetic energy; $\alpha > 0$ is appropriate for boson hopping $t_{rr'} < 0$.  In Eq.~(\ref{eqn:uud_perm}), column $j$ of the Permanent matrix is given by the $j$-th orbital (centered on $A_j$) evaluated on the occupied sites $\{r_i\}$.\cite{Ceperley78, Handbook}  One can loosely connect this wavefunction with a picture starting from the ``Ising'' limit, $J^z \gg \vert t \vert$, and perturbatively building in boson kinetic energy effects.\cite{Honecker99}

Table~\ref{table:uud} compares the trial energies for different number of variational parameters.  Excluding any Jastrow factor, the single-parameter trial state already captures the important exchange energies; for example, it is closer to the exact ground state in the zero momentum sector than to the first excited state in this sector (not shown).\cite{Bernu94}  Adding a short range Jastrow factor further improves the trial energy, while longer range Jastrow parameters are unimportant since correlations in the Mott insulator decay rapidly (see Fig.~\ref{fig:correlation} in Appendix~\ref{app:corr}).  We see that the simplest localized orbitals extending only to nearest neighbour sites (and with relatively small amplitude $\alpha \sim 0.23$) perform very well, which suggests strong $uud$ order in the 1/3-filled system.\cite{Chubukov91, Honecker99}  Indeed, for the optimal wavefunction, we calculate the boson density to be 0.76 on $A$ sites and 0.12 on $BC$ sites.

\subsection{$Y$ state at $n\gtrsim 1/3$}
\label{subsec:Y}

\begin{table}
  \begin{tabular*}{\columnwidth}{@{\extracolsep{\fill}}cccccc}
    \hline
    Trial state & $\mathrm{N_{par}^{orb}}$ & $\mathrm{N_{short}^{Jas}}$ & $\mathrm{N_{long}^{Jas}}$ & $\mathrm{N_{par}^{tot}}$ & E/bond\\
    \hline
    classical &   &   &   &   & -0.0961\\
    spiral    & 0 & 1 & 2 & 4 & -0.1424\\
    $Y$       & 1 & 0 & 0 & 1 & -0.1461\\
              & 1 & 1 & 0 & 2 & -0.1477\\
              & 1 & 2 & 0 & 3 & -0.1478\\
    exact     &   &   &   &   & -0.1489\\
    \hline
  \end{tabular*}
  \caption{Comparison of $Y$ trial energies for different number of short-range Jastrow parameters for 13 bosons on a $6\times 6$ cluster. The $Y$ state is constructed by adding one boson to the $uud$ state as described in the text; we allowed the same variational parameters as in the $uud$ case in Table~\ref{table:uud}. We also show trial energy for the spiral state with 4 variational parameters, which has higher energy than the $Y$ state.}
  \label{table:Y}
\end{table}

Starting from the $uud$ wavefunction where we have good exchange energies between sublattices $A$ and $BC$, we construct a candidate for the \textit{Y} supersolid phase by adding bosons to an extended orbital on $BC$:
\begin{eqnarray}
|\psi_Y \ra &=& \left( \sum_r \phi_{BC}^{\rm ext}(r) b_r^\dagger \right)^{N_b - N/3} |\psi_{uud} \ra \label{Y} ~, \\
\phi_{BC}^{\rm ext}(r) &=& \left\{
  \begin{array}{ll}
    +1,& r\in B\\
    -1,& r\in C\\
     0,& r\in A
  \end{array}\right.
\end{eqnarray}
Just as in the $uud$ case, the wavefunction can be written as a $N_b \times N_b$ permanent.  The first $N/3$ columns contain the same $\phi_j^{\rm loc}$ orbitals as in the $uud$ state, while the remaining $N_b - N/3$ columns all contain the extended orbital $\phi_{BC}^{\rm ext}$ residing on the $B$ and $C$ sublattices.  The alternating signs of the extended orbital are appropriate for bosons hopping on the $BC$ honeycomb with $t_{rr'} < 0$.  Nearest neighbour contacts on $BC$ are suppressed by adding a Jastrow factor.  

Table~\ref{table:Y} compares the $Y$ energy against spiral and ED energies for 13 bosons on the $6 \times 6$ cluster.  Our $Y$ state is close to the ED ground state from Bernu\etal;\cite{Bernu94} thus, the trial energy is below the first excited state in the same sector (not shown), while the spiral is significantly higher.

We consider such $Y$ states for all boson densities above $1/3$ and find them to give lowest trial energies among all our states for $N_b = 13, \dots, 15$.  We discuss properties of the $Y$ states in Appendix~\ref{app:corr}.  Here we note an interesting feature that boson correlations are long-ranged for $B$ and $C$ sublattice sites but are short-ranged for $A$ sites.  The $A$ sublattice remains ``Mott-insulating'' despite the superfluid on the $BC$ honeycomb.  The absence of the ``proximity effect'' on the $A$ sublattice is due to cancellations from alternating superfluid order parameter on the $B$ and $C$ sublattices.\cite{Wang09}  In particular, just as in the $uud$ case, we cannot construct Bijl-Jastrow-type wavefunction for the $Y$-state.

\subsection{Spiral state at $n\lesssim 1/2$}
\label{subsec:spiral}

\begin{table}
  \begin{tabular*}{\columnwidth}{@{\extracolsep{\fill}}cccccc}
    \hline
    Trial state & $\mathrm{N_{par}^{orb}}$ & $\mathrm{N_{short}^{Jas}}$ & $\mathrm{N_{long}^{Jas}}$ & $\mathrm{N_{par}^{tot}}$ & E/bond\\
    \hline
    classical &   &   &   &   & -0.1250\\
    $Y$       & 1 & 1 & 2 & 4 & -0.1774\\
    spiral    & 0 & 1 & 0 & 1 & -0.1728\\
              & 0 & 1 & 0 & 2 & -0.1791\\
              & 0 & 1 & 2 & 4 & -0.1795\\
    exact     &   &   &   &   & -0.1868\\
    \hline
  \end{tabular*}
  \caption{Comparison of spiral energies for different number of variational parameters for 18 bosons on a 36-site cluster. The extended orbital of the spiral state has amplitudes $1$, $e^{i2\pi/3}$, and $e^{i4\pi/3}$ for sites on sublattices $A$, $B$, and $C$ respectively. The first spiral case has one nnb Jastrow pseudopotential which is taken to be the same between any pair of nnb sites, while the second spiral case has an additional parameter $\gamma$ for the Huse-Elser phase factor. The third case has two more parameters $w$ and $p$ for a long range pseudopotential $\frac{w}{\vert i-j\vert^p}$ between any pair of sites $i$ and $j$. We also show trial energy for the $Y$ state with 4 variational parameters, which has slightly higher energy than the spiral state.}
  \label{table:spiral}
\end{table}

At half-filling, the $120^\circ$ magnetically ordered state (spiral) is believed to be the ground state.  We use Huse and Elser wavefunction,\cite{Huse88} which generalizes Bijl-Jastrow-type wavefunction by including complex 3-body terms, to accurately describe the corresponding superfluid state of bosons near half-filling.  In this wavefunction, all the bosons reside on an extended orbital,
\begin{eqnarray}
  |\psi_S \ra = e^{i\sum_{ijk} \gamma_{ijk}\ n_i n_j n_k} 
\left( \sum_r e^{i \vec{Q}\cdot \vec{r}}\ b_r^\dagger \right)^{N_b} |0 \ra, 
\label{spiral}
\end{eqnarray}
with $\vec{Q} = (4\pi/3,0)$.
Despite frustration, the bosons gain some kinetic energy while hopping along any lattice link.  Nearest neighbour contacts are suppressed by adding a long range Jastrow factor.  The three-body phase factor, which respects the symmetries of the classical state, serves as an additional variational parameter.  For details, the reader is referred to the original Ref.~\onlinecite{Huse88}.

Among our trial states, the Huse-Elser wavefunction has lower energy than the $Y$ state for $N_b = 16, \dots, 18$, but only by a very small amount (cf.\ Fig.~\ref{fig:isotropic_energies}).  Table~\ref{table:spiral} shows that the 18-boson spiral energy is only slightly lower than the $Y$ energy.  This is perhaps not surprising since the classical $120^\circ$ order may be viewed as the spiral or $Y$-shape order depending on the plane's orientation.  A recent variational study\cite{Heidarian09b} using different constructions of the spiral and $Y$ states obtained $-0.1827$ for their many-parameter spiral state, which is also lower than their $Y$ trial energy by a small amount similar to that in our study.  Other recent works\cite{Wang09, Jiang09} observed an abrupt change from the spiral to $Y$ supersolid in the half-filled model as the spin anisotropy is varied through the SU(2)-invariant Heisenberg point.  In principle, thinking in terms of wavefunctions, the spiral and $Y$ can be distinct phases with different postulated symmetry breaking also in the SU(2)-invariant model.  However, this could also be a plane reorientation transition, and the closeness in energy of the $Y$ trial states reflecting their ability to capture the $120^\circ$ spiral order.

\subsection{$V$ state at $n\lesssim 1/3$}
\label{subsec:Vperm}

\begin{table}
  \begin{tabular*}{\columnwidth}{@{\extracolsep{\fill}}cccccc}
    \hline
    Trial state & $\mathrm{N_{par}^{orb}}$ & $\mathrm{N_{short}^{Jas}}$ & $\mathrm{N_{long}^{Jas}}$ & $\mathrm{N_{par}^{tot}}$ & E/bond\\
    \hline
    classical         &   &   &   &   & -0.0683\\
    spiral            & 0 & 1 & 2 & 4 & -0.1081\\
    $V_\mathrm{perm}$ & 1 & 0 & 0 & 1 & -0.1141\\
                      & 1 & 1 & 0 & 2 & -0.1153\\
                      & 1 & 2 & 0 & 3 & -0.1154\\
    exact             &   &   &   &   & -0.1161\\
    \hline
  \end{tabular*}
  \caption{
Comparison of $V_{\rm perm}$ energies for different number of variational parameters for 11 bosons on a 36-site cluster. The $V_{\rm perm}$ state is constructed by removing one boson from the $uud$ state as described in the text, and the parameters here are of the same type as in the $uud$ case in Table I. We also show trial energy for the spiral state with 4 variational parameters, which has higher energy than the $V_{\rm perm}$ state.
}
  \label{table:Vperm}
\end{table}

Let us now consider densities slightly less than 1/3.  We start from the $uud$ state and picture it as a filled $A$ sublattice.  An appealing scenario is to introduce holes and let them move around on $A$ and condense.  We automatically retain charge order selecting the $A$ sublattice vs $B$ and $C$.  The condensation of holes on the $A$ gives boson superfluid order there and by proximity effect also on the $B$ and $C$ sublattices.  Since $t_{AB} = t_{AC} < 0$, we expect the phase angle on the $BC$ to be shifted by $\pi$ from the $A$.  The resulting supersolid is precisely the $V$ state.

Direct wavefunction implementation of this scenario is described in Appendix~\ref{app:Vperm} and leads to a sum of permanents, which becomes prohibitively costly to evaluate for more than few holes.  In the appendix, we also motivate a qualitatively similar wavefunction with a simpler amplitude given by a single permanent,
\begin{eqnarray}
  \la \left\lbrace r_k\right\rbrace\vert \psi_{V_{\rm perm}}\ra =
  \mathrm{Perm}
  \begin{pmatrix}
    \phi_1^\mathrm{loc}(r_1) & \dots & \phi_{N/3}^\mathrm{loc}(r_1) \\
    \vdots & \ddots & \vdots \\
    \phi_1^\mathrm{loc}(r_{N_b}) & \dots & \phi_{N/3}^\mathrm{loc}(r_{N_b}) \\
    1 & \dots & 1 \\
    \vdots & \ddots & \vdots \\
    1 & \dots & 1
  \end{pmatrix}.\label{eqn:Vperm_matrix}
\end{eqnarray}
The first $N_b$ rows contain the localized orbitals of the $uud$ construction evaluated at the boson positions, while the remaining $N/3 - N_b$ rows are filled with $1$-s corresponding to ``zero wavevector'' condensate of holes (see Appendix~\ref{app:Vperm} for details).

Table~\ref{table:Vperm} shows the $V_{\rm perm}$ energy for 11 bosons on the $6 \times 6$ cluster.  Being a descendant of the excellent $uud$ state, even with no Jastrow factor the $V_{\rm perm}$ performs very well and lies roughly half-way between the ground state and the first excited state with the same quantum numbers (the latter is not listed in the table).  In particular, the $V_{\rm perm}$ clearly wins over the spiral superfluid with uniform density.  Just as in the $uud$ case, adding simple short-range Jastrow parameters further improves the trial energy of the $V_{\rm perm}$ state.  At this stage, we did not include long-range Jastrow pseudopotentials, which would be needed for correct long-wavelength description\cite{Chester67} of superfluid correlations in the $V$ phase.

The $V_{\rm perm}$ state gives our best variational energies for $N_b = 6, \dots, 11$.  In Appendix~\ref{app:corr}, we measure properties of this state and verify the superfluid order with opposite signs on the $A$ and $BC$ sublattices as anticipated above.

\subsection{$V$ state at $n\ll 1/3$}
\label{subsec:VBJ}

\begin{table}
  \begin{tabular*}{\columnwidth}{@{\extracolsep{\fill}}cccccc}
    \hline
    Trial state & $\mathrm{N_{par}^{orb}}$ & $\mathrm{N_{short}^{Jas}}$ & $\mathrm{N_{long}^{Jas}}$ & $\mathrm{N_{par}^{tot}}$ & E/bond\\
    \hline
    classical        &   &   &   &   & 0.13542\\
    spiral           & 0 & 1 & 2 & 4 & 0.12885\\
    $V_\mathrm{BJ}$   & 1 & 0 & 0 & 1 & 0.13216\\
                     & 1 & 1 & 0 & 2 & 0.12913\\
                     & 1 & 1 & 2 & 4 & 0.12869\\
    exact            &   &   &   &   & 0.12845\\
    \hline
  \end{tabular*}
  \caption{Comparison of $V_{\rm Bijl-Jastrow}$ trial energies for different number of variational parameters for 3 bosons on a 36-site cluster. The extended orbital of the $V_{\rm Bijl-Jastrow}$ state has amplitude $e^{\mu/2}$ on sublattice $A$ and $-1$ on sublattices $B$ and $C$; $\mu$ is the only variational parameter in the first listed $V_{\rm Bijl-Jastrow}$ case. The second $V_{\rm Bijl-Jastrow}$ case has one nnb Jastrow pseudopotential which is taken to be the same between any pair of nnb sites, while the third case has two additional parameters for a long range pseudopotential which is the same as in the spiral case in Table~\ref{table:spiral}. We also show trial energy for the spiral state with 4 variational parameters, which has slightly higher energy than the $V_{\rm Bijl-Jastrow}$ state.}
  \label{table:Vhuse}
\end{table}

The above wavefunction for the $V$ phase is obtained from the strong $uud$ state and a priori is not expected to remain good at low density.  Here we consider an alternative construction of the $V$ supersolid using Bijl-Jastrow-type wavefunction, working directly with bosons and condensing them into an appropriate extended orbital,
\begin{eqnarray}
|\psi_{V, {\rm Bijl-Jastrow}} \ra &=& 
\left( \sum_r \phi_V^{\rm ext}(r)\ b_r^\dagger \right)^{N_b} |0 \ra 
\label{V_Bijl-Jastrow}\\
\phi_V^{\rm ext}(r) &=& \left\{
\begin{array}{l}
  e^{\mu/2},\ r\in A\\
  -1,\ r\in B, C
\end{array}\right.\label{eqn:phiV}
\end{eqnarray}
This orbital has opposite signs on the $A$ and $BC$ sublattices as expected from Fig.~\ref{fig:boson_orderings}d.  The ``chemical potential'' $\mu$ on the $A$ sublattice allows us to control the charge order.  Similar to other wavefunctions for states with superfluid order, it is necessary to include a long-range Jastrow pseudopotential.

As we argue below, this wavefunction is a natural candidate at low boson densities.  On the $6 \times 6$ cluster, it optimizes better than the $V_{\rm perm}$ state for $N_b \leq 6$ and also has better energy than the spiral state, see Fig.~\ref{fig:isotropic_energies}.  As an example of variational results, Table~\ref{table:Vhuse} shows the $V_{\rm Bijl-Jastrow}$ energy for 3 bosons on the $6 \times 6$ cluster.  We see that the $V$ state is slightly better than the spiral state.  However, both states are quite close in energy and close to the exact ground state.  We discuss this more below and see what we can infer about the competition between the coplanar and spiral states from ED spectroscopy.

First, we want to connect the competing $V_{\rm Bijl-Jastrow}$ and spiral states with physics at low boson densities.  In the absence of interaction, the kinetic energy minimizes at two distinct points in the Brillouin zone, $\pm\vec{Q} = \pm(4\pi/3,~0)$.  Boson condensation at one point gives rise to the spiral phase; schematically, the spiral wavefunction is given by $(b_{\vec{Q}}^\dagger)^{N_b} |0\ra$ [or $(b_{-\vec{Q}}^\dagger)^{N_b} |0\ra$ for the opposite wavevector].  A more complex condensation pattern including both points produces a coplanar state, with schematic wavefunction $(e^{i\alpha} b_{\vec{Q}}^\dagger + \hc)^{N_b} |0\ra$.  When $\alpha=0$, this gives boson orbital $\phi(r) = \cos(\vec{Q}\cdot \vec{r})$ taking values $\{+1, -1/2, -1/2\}$ on the three sublattices, which is essentially the $\phi_V^{\rm ext}$ orbital in Eq.~(\ref{eqn:phiV}).  On the other hand, $\alpha = \pi/2$ corresponds to a different state with zero boson density on one sublattice and alternating superfluid phases on the other two sublattices; in terms of spins, this is a coplanar ``$\Psi$''-type state which has similar symmetry to the $Y$ state in Fig.\ref{fig:spin_orderings}b, but with the vertical spin flipped up.  For either $V$ or $\Psi$, there are two more degenerate states given by lattice translations or equivalently by adding $\pm 2\pi/3$ to the phase $\alpha$.  Reference \onlinecite{Nikuni95} studies the dilute boson problem analytically for the isotropic lattice and predicts that four-boson interactions select coplanar states; this study does not resolve between the $V$ and $\Psi$ states, which requires considering six-boson terms.

Returning to our example with 3 bosons on the $6 \times 6$ lattice, the ED ground state has momentum quantum number $\vec{k} = \vec{Q}$ (there is also a degenerate state with opposite momentum), while we also find two very close states with $\vec{k} = 0$, one with even and the other with odd parity under inversion.  This can be traced to 4 degenerate eigenstates of the kinetic energy, 
\begin{equation}
\left\{ (b_{\vec{Q}}^\dagger)^3, \quad (b_{-\vec{Q}}^\dagger)^3, \quad
(b_{\vec{Q}}^\dagger)^2 b_{-\vec{Q}}^\dagger, \quad b_{\vec{Q}}^\dagger (b_{-\vec{Q}}^\dagger)^2
\right\} ~.
\label{2boson}
\end{equation}
Our spiral wavefunction construction gives essentially the first two states with momentum quantum number $\vec{k}=0$.  Our 3 degenerate $V$-states, upon making combinations that are momentum eigenstates, give an even-parity $\vec{k}=0$ combination as well as the last two states with $\vec{k} = \pm \vec{Q}$ from Eq.~(\ref{2boson}).  Finally, the 3 degenerate $\Psi$-type states give an odd-parity $\vec{k} = 0$ combination and the same two $\vec{k} = \pm \vec{Q}$ states.  It is clear that these trial states are not independent for this small number of bosons; we cannot resolve the phases, but we can start looking for some tendencies.  For example, we can view the fact that the $\vec{k} = \pm \vec{Q}$ are lower in energy than $\vec{k} = 0$ as an indication for the coplanar states being better than the spiral.  In principle, we could also try to resolve between the $V$ and $\Psi$ by comparing the even/odd-parity $\vec{k}=0$ states, but the splitting is too tiny.

We have similarly examined ED spectra with $N_b = 4, \dots, 9$ bosons, paying attention to near degeneracy of ground states and their quantum numbers.  The resolution between the spiral and coplanar states due to interactions becomes clearer with increasing density, and in each instance the ED data is consistent with the coplanar states being better, which is also supported by the VMC data.  As far as the resolution between the $V$ and $\Psi$ states is concerned, we can not tell anything with boson number below $6$, while for higher boson density we start seeing evidence in favor of the $V$ state.  The $V$ state is expected coming from the $n=1/3$ plateau as we discussed earlier. One possibility is that $V$ occurs for all $n<1/3$, but we cannot rule out transition to the $\Psi$ coplanar state at low densities.  Our VMC study of the spatially anisotropic model on larger clusters in Sec.~\ref{subsec:VvsS} also suggests that the coplanar phase (incommensurate in this case, so there is no distinction between $V$ and $\Psi$) wins over the spiral also for a range of anisotropies, strengthening the conclusions here on the coplanar versus spiral energetics.

\subsection{Summary of trial energies on the isotropic lattice}

\begin{figure}[t] 
  \centering
  \includegraphics[width=\columnwidth]{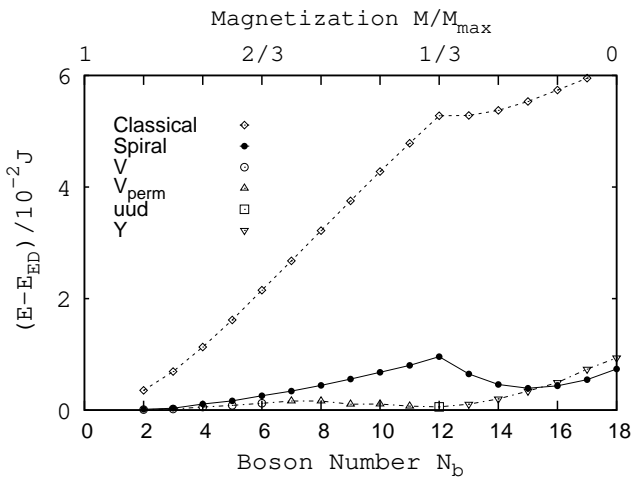}
  \caption{Comparison of variational energies (per bond) for the 36-site cluster with isotropic exchanges.  The trial energies for the planar states are obtained using Bijl-Jastrow-type wavefunction for $V$ ($N_b \leq 6$) and permanent-type wavefunctions for $V$ ($7\leq N_b\le 11$), $uud$ ($N_b=12$), and $Y$ ($N_b \geq 12$).  The spiral trial energies are obtained using the Huse-Elser wavefunction.  For clarity, ED ground state energies are subtracted at respective boson numbers.  The classical energy curve provides a reference for judging stabilization of specific phases by quantum fluctuations.}
  \label{fig:isotropic_energies}
\end{figure}

Fig.~\ref{fig:isotropic_energies} summarizes the spin energies (per bond) of competing trial states calculated for the $6\times 6$ cluster with periodic boundary conditions for all $N_b$.  For a better comparison of the accuracies of these trial wavefunctions, we subtract the ED ground state energy at each boson density.  The energies of the classical state are included to emphasize the stabilization of specific phases by quantum fluctuations.  Our wavefunctions are particularly accurate in the vicinity of the plateau, and also at low boson densities (higher fields). In the latter regime, the classical energies approach ED values at low densities, indicating vanishing quantum fluctuations.\cite{Chubukov91, Nikuni95, Veillette06}

\subsection{Magnetization process on the isotropic lattice}

\begin{figure}
  \centering
  \includegraphics[width=\columnwidth]{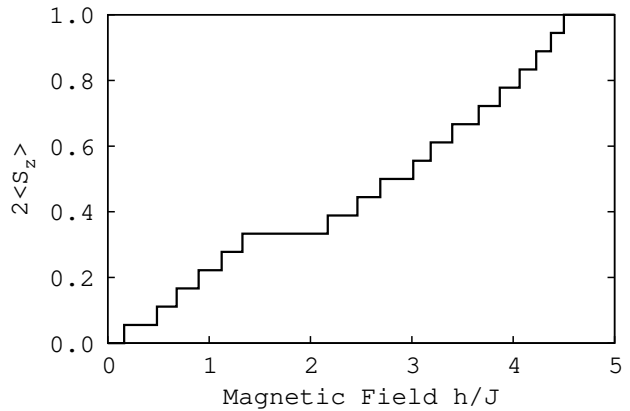}
  \caption{Magnetization curve of the 36-site cluster obtained using the variational energies.  The pronounced $uud$ plateau agrees well with the ED results in Ref.~\onlinecite{Honecker04}. This shows that the variational tool is able to capture this Mott-insulator as well as nearby supersolid phases.
}
  \label{fig:isotropic_mag}
\end{figure}

Using the trial energies from our studies at fixed $N_b$, we can work out the magnetization curve as a function of field $h$. Fig.~\ref{fig:isotropic_mag} shows this for the 36-site cluster.  The boundaries of the $1/3$ magnetization plateau are determined by the energy gaps to adding or removing one boson to the $uud$ state.  Since our permanent constructions give very good trial $Y$ and $V$ states in this regime, the estimate of the plateau range is quite accurate.  To check finite size effects, we repeated the calculation on a 63-site cluster and obtained critical fields $H_{c1}\approx 1.4 J$ and $H_{c2}\approx 2.2 J$.

\subsection{2-parton trial wavefunctions and 
alternative construction of $uud$ state at $n=1/3$}\label{subsec:2p}

The above direct study using spiral, $Y$, $uud$, and $V$ states is sufficient to describe the phase diagram of the spatially isotropic triangular antiferromagnet in the field.
We now consider a versatile set of trial wavefunctions which we will call ``2-parton'' states.  One motivation is to give a practical realization of the Chern-Simons flux attachment treatment in Ref.~\onlinecite{Misguich01}.
(The relation between the parton and Chern-Simons approaches is discussed in Ref.~\onlinecite{DBL} and citations therein.)  Another motivation is to prepare for an anisotropic lattice study in Sec.~\ref{sec:anisotropic}.  
We should say from the outset that while such parton construction is typically used to produce fractionalized (spin liquid) states, it can also be used to give more conventional states such as CDW of bosons with no topological order as discussed below.

We represent the boson operator in terms of two fermions, $b = d^{(1)} d^{(2)}$, subject to constraint $b^\dagger b = d^{(1)\dagger} d^{(1)} = d^{(2)\dagger} d^{(2)}$ on each site.  Imagine some ``mean field Hamiltonian'' for each parton flavor,
\begin{eqnarray}
\hat{H}^{(n)}_{\rm mf} &=& -\sum_{\langle rr' \rangle} 
\left( |t_{rr'}^{(n)}| e^{i a_{rr'}^{(n)}} d_r^{(n)\dagger} d_{r'}^{(n)} + \hc
\right) \nonumber\\
&& -\sum_r \mu_r^{(n)} d_r^{(n)\dagger} d_r^{(n)} ~.
\label{Hp}
\end{eqnarray}
Here we write the parton hopping amplitudes (which can be complex) as $t_{rr'}^{(n)} = |t_{rr'}^{(n)}| e^{i a_{rr'}^{(n)}}$; we also allow site-dependent chemical potentials $\mu_r^{(n)}$ to test CDW tendencies.  The $t_{rr'}^{(n)}$ and $\mu_r^{(n)}$ are variational parameters.  We solve $\hat{H}^{(n)}_{\rm mf}$ and fill up the corresponding Fermi seas with $N_{d_1} = N_{d_2} = N_b$ particles.  A valid bosonic wavefunction is obtained by applying a Gutzwiller projection such that every site is either empty ($n_b=n_{d_1}=n_{d_2}=0$) or contains both $d_1$ and $d_2$ partons ($n_b=n_{d_1}=n_{d_2}=1$):
\begin{eqnarray}
  \vert \psi_{2p} \ra = \hat{P}_G \prod_{q_n\in FS_n} d_{q_1}^{(1)\dagger} d_{q_2}^{(2)\dagger} \vert 0 \ra.
\end{eqnarray}
For each boson configuration, the amplitude is given by a product of two Slater determinants.  One feature of this construction follows from the fermionic statistics which provides an inherent repulsive Jastrow effect for the particles.  This effect can be tuned as follows,
\begin{eqnarray}
  \la \left\lbrace r_k\right\rbrace \vert \psi_{2p}\ra 
= {\rm det}_1 \cdot {\rm det}_2 \cdot \vert {\rm det}_1 \vert^{p_1 - 1} \cdot \vert {\rm det}_2 \vert^{p_2 - 1} ~,
\label{det1det2}
\end{eqnarray}
which preserves the ``sign structure'' of the wavefunction while allowing more variational freedom with parameters $p_1$ and $p_2$.  Numerical calculations can be performed using efficient determinantal Monte Carlo techniques.\cite{Ceperley77}

Besides treating boson repulsion, we want to have good kinetic energy.  We can write the frustrated boson hopping amplitudes in Eq.~(\ref{eqn:frustration}) as $t_{rr'}^{(b)} = |t_{rr'}^{(b)}| e^{i a_{rr'}^{(b)}}$ and view this as a problem in an external orbital field producing flux $\pi$ through each triangle.\cite{KalmeyerLaughlin89}  To capture this in the parton treatment, we view $d^{(1)}$ and $d^{(2)}$ as charged particles whose charges add up to that of the boson $b$; we therefore require
\begin{eqnarray}
e^{i a_{rr'}^{(1)}}  e^{i a_{rr'}^{(2)}} =  e^{i a_{rr'}^{(b)}} .
\label{t1t2}
\end{eqnarray}
Thus the parton mean field Hamiltonian should contain fluxes such that for the two flavors they add up to the original flux seen by the bosons.  We can still make different choices, say, for the $d_1$; however, once the $a_{rr'}^{(1)}$ are fixed, then the $a_{rr'}^{(2)}$ are uniquely determined.  

We first discuss what we will call ``Chern-Simons'' states that realize the idea in Ref.~\onlinecite{Misguich01}.  For the $d_1$ hopping, we take uniform flux of $n \pi$ per triangle, where $n$ is the boson density per site.  With this choice, the ${\rm det}_1$ Slater determinant fills the ``lowest Landau level'' band and gives a finite lattice version of the usual Chern-Simons factor $\prod_{i<j} (z_i - z_j)$.  We can loosely view the ${\rm det}_1$ as performing flux attachment transformation from the bosons to the $d_2$ fermions.\cite{DBL} Upon subsequent ``flux smearing'' mean field, the $d_2$ see flux $(1-n)\pi$ per triangle.  In the absence of site-dependent chemical potentials and for some rational densities, the ${\rm det}_2$ Slater determinant is gapped, and the boson wavefunction realizes a fractionalized ``chiral spin liquid''.\cite{Misguich01, KalmeyerLaughlin89, WenWilczekZee89}  We have tried these ``topological'' states for several densities such as $n = 1/3, 1/4, 1/6$ on the isotropic triangular lattice and found that they are poor compared with the $uud$ and $V$ states described earlier.  Thus the interesting proposal of plateaus due to chiral spin liquid states is not realized on this lattice.\cite{Misguich01}

We now specialize to density $n = 1/3$ and allow a chemical potential on the $A$ sublattice: $\mu_A \neq 0, \mu_B = \mu_C = 0$.  We find that optimal $\mu_A^{(1)}, \mu_A^{(2)}$ are large and produce strong CDW order in the mean field state.  When this happens, the trial boson state Eq.~(\ref{det1det2}) is no longer topological in nature.  Indeed, if the parton hopping is set to zero, this construction simply gives the classical $\sqrt{3} \times \sqrt{3}$ CDW state.  The particles completely occupy sublattice $A$, and there is a large gap at the parton Fermi level.  Adding small hopping does not close this gap but only builds in some charge fluctuations into the parton mean field and thus into the boson trial state.  Working perturbatively in $t^{(n)}/\mu^{(n)}_A$, the leading modification to the classical boson CDW wavefunction is to add configurations where one particle moves from a site $A_j$ to a neighbor $r$.  The amplitude for such a configuration is proportional to 
$t^{(1)}_{A_j, r} t^{(2)}_{A_j, r} / (\mu^{(1)}_A \mu^{(2)}_A) 
\sim t^{(b)}_{A_j, r} / \mu^{(b)}_A$, where we have kept track of all signs and introduced schematically boson charge gap $\mu^{(b)}_A$.
The result is similar to the perturbative picture of the CDW working directly in the boson language that motivated the wavefunction Eq.~(\ref{eqn:uud_wf}).  Thus at this level the 2-parton states with strong CDW potential are qualitatively the same as the permanent $uud$ state in Sec.~\ref{subsec:uud}.

The above leading order structure holds for all 2-parton states satisfying Eq.~(\ref{t1t2}).  At higher order, the states will differ, and amplitudes can be complex in general: e.g., the Chern-Simons wavefunction described above is complex-valued.  On the other hand, the permanent $uud$ wavefunction is real.  The boson Hamiltonian Eq.~(\ref{eqn:Hb}) is invariant under complex conjugation in the number basis, and the $uud$ state preserves this symmetry.  We can construct a real-valued 2-parton state by taking the $d_1$ fluxes to be $0$ or $\pi$ through up or down triangles; the $d_2$ partons see correspondingly $\pi$ and $0$ fluxes.  We will call this state ``$U1B$''.  It was originally discussed at half-filling in Ref.~\onlinecite{ZhouWen}, where (in the absence of chemical potentials) it has Dirac nodes at the Fermi level and realizes so-called Algebraic Spin Liquid state.  This particular state has a good trial energy in the Heisenberg model,\cite{Yunoki06, DBL, Heidarian09b} and can be viewed as a more elaborate real-valued version of the Laughlin-Kalmeyer state (see Sec.~IIC of Ref.~\onlinecite{DBL} for more discussion).  Away from half-filling, the $U1B$ mean field state has Fermi surfaces of partons and may be unstable to a mechanism described in Ref.~\onlinecite{Ran09}.   However, this is not a direct concern here since we are gapping out the state by adding large $\mu_A$ potential and are connecting to the strong CDW of bosons.  The virtue of using the 2-parton framework is that it naturally builds in small charge fluctuations as described above, and determinants are easier to compute as opposed to permanents.   Using this construction for the isotropic $6\times 6$ lattice at $n=1/3$, we obtain a very competitive energy $-0.1341$ (cf.\ trial energies in ~Table~\ref{table:uud}) with $\mu_A \approx 2$ and $p \approx 0.75$.  (We also obtain close trial energy using the Chern-Simons state with strong CDW potential, in agreement with the earlier discussion that all 2-parton states can similarly capture leading local charge fluctuations when the charge order is strong).  The 2-parton constructions are particularly useful on the anisotropic lattice to be discussed in Sec.~\ref{sec:anisotropic}, since they naturally connect to the decoupled chains limit and allow us to detect where the quasi-1d physics sets in and explore CDW instabilities.


\section{Anisotropic triangular antiferromagnet}\label{sec:anisotropic}
Motivated by the unknown phases of $\textrm{Cs}_2\textrm{CuBr}_4$ in the field,\cite{Fortune09} we extend our study to the spatially anisotropic lattice.  As the phase diagram is much more complex, it is appropriate to begin this section with a short review of the different regimes and phases discussed in theoretical literature.  Next, we describe anisotropic extensions of the wavefunctions introduced in Sec.~\ref{sec:isotropic} and then present our variational results.

\subsection{Review of phases from theoretical studies}
In this review, it is convenient to refer to a schematic phase diagram shown in Fig.~\ref{fig:phase_diagram3}, where we parametrize the anisotropy using $\delta=1-J'/J$.

In zero magnetic field (bottom axis in Fig.~\ref{fig:phase_diagram3}), variational studies suggest that the spiral phase remains stable for small lattice anisotropy, while the more anisotropic region may contain one or two spin liquid phases.\cite{Huse88,Yunoki06,Heidarian09a} This is supported by an ED/DMRG study which found signatures of spin liquid for $J'/J < 0.78$ from numerical measurements of spin structure factor, excitation energy gap, and spin correlation.\cite{Weng06} However, for large anisotropy, an analytical study near the decoupled chains limit predicts a collinear antiferromagnetic order.\cite{Starykh07} This suggests that the zero field limit is a challenging region which remains unsettled.

In the high field limit near full polarization (top phase boundary in Fig.~\ref{fig:phase_diagram3}), analytical studies of the dilute boson gas show that the $V$ phase (commensurate and incommensurate) is the likely candidate near the spatially isotropic regime,\cite{Nikuni95} while the spiral phase dominates for strong anisotropy.\cite{Veillette06}

At intermediate fields, an interacting spin wave expansion about the $uud$ plateau in large $S$ and low anisotropy limit (left axis in Fig.~\ref{fig:phase_diagram3}) shows that the plateau extends considerably into the anisotropic region, with commensurate coplanar $Y$ and $V$ phases present next to the plateau.\cite{Alicea09}  In addition, incommensurate coplanar and distorted spiral phases are also predicted for larger anisotropy.  

In the nearly decoupled chains limit (right axis in Fig.~\ref{fig:phase_diagram3}), Ref.~\onlinecite{Starykh07} argues that the interchain coupling is a relevant perturbation and can induce various boson CDW phases or a spiral phase.  The former happens for small and intermediate fields, while the latter is expected near the saturation field.

\subsection{Anisotropic versions of wavefunctions}

Although our study began with a goal of constructing accurate wavefunctions for identifying the unknown phases of $\textrm{Cs}_2\textrm{CuBr}_4$, it quickly became clear that this is far from easy.  The number of theoretically possible phases reviewed above is already very rich, with different physics regimes requiring different mindsets.  Nevertheless, the variational approach is a useful tool for obtaining quantitative insights into the energetics of various phases, since it applies directly to the spin-1/2 problem at hand and goes beyond approximate treatments like large-$S$ and mean field.  Encouraged by our success for the isotropic problem in the field, we apply this tool to the anisotropic case, while being critical of the limitations of the variational approach.

We consider the anisotropic triangular lattice antiferromagnet with $J'/J \leq 1$, where $J$ and $J'$ are the coupling constants of horizontal and oblique nearest neighbour links (see Fig.~\ref{fig:ani_lattice}).  The spatially anisotropic wavefunctions used in this section contain appropriate modifications to the wavefunctions in Sec.~\ref{sec:isotropic}. 

\textit{Permanent constructions}:  For the wavefunctions from Secs.~\ref{subsec:uud}, \ref{subsec:Y}, and \ref{subsec:Vperm}, the localized orbitals used in the $uud$, $Y$, and $V_{\rm perm}$ now include an additional parameter $\alpha'$:
\begin{eqnarray}
  \phi_j^{\rm loc}(r) &=& \left\{
  \begin{array}{ll}
    1, & r = A_j\\
    -\alpha, & r = \text{horizontal n.n.\ of $A_j$;}\\
    -\alpha', & r = \text{oblique n.n.\ of $A_j$;}\\
    0, & \text{otherwise.}
  \end{array}
  \right.
\end{eqnarray}

\textit{Spiral}: Our treatment of the spiral requires separate discussion. The $120^\circ$ spiral generalizes to an incommensurate spiral with wavevector $\vec{Q}$.\cite{Veillette06,Alicea09}  However, in a finite sample, periodic boundary conditions would bias against the incommensurate order.  We can mitigate this effect by considering appropriate phase twists at the boundaries that accommodate such $\vec{Q}$.  For computations, it is convenient to perform a gauge transformation that spreads the twist uniformly across the sample; the resulting Hamiltonian is then translationally invariant:
\begin{eqnarray}
  \hat{H}_{\rm twisted} = -\sum_{\la rr'\ra} 
\left(t_{rr'} e^{i \vec{Q} \cdot \vec{e}_{rr'}} b_r^\dagger b_{r'} + \hc \right) + H_{\rm int} ~,
\label{Htwisted}
\end{eqnarray}
where $\vec{e}_{rr'}$ is the displacement vector from $r$ to $r'$.  The twisted Hamiltonian is used only for calculating the incommensurate spiral energies while all other trial energies are evaluated using the original Hamiltonian with no twist.

\textit{Jastrow factors}: To accommodate spatial anisotropy, we introduce additional parameters into the nearest neighbour and long-range pseudo-potentials as follows:
\begin{eqnarray}
u(r,r') =
\left\{
\begin{array}{l}
  \begin{array}{ll}
    w,  & \quad\text{if $r$ and $r'$ are horizontal n.n.}\\
    w', & \quad\text{if $r$ and $r'$ are oblique n.n.}
  \end{array}\\
  \frac{A}{[\alpha^2 (x - x')^2 + (y - y')^2]^{p/2}},
  \quad\text{otherwise.}
  \end{array}
\right.\label{Eq:ani_pseudopotential}
\end{eqnarray}

\textit{2-parton}: To obtain spatially anisotropic versions of the 2-parton wavefunctions from Sec.~\ref{subsec:2p}, we allow the mean field hopping amplitudes in Eq.~(\ref{Hp}) to be anisotropic:
\begin{eqnarray}
  t_{rr'}^{(n)} = \left\{
  \begin{array}{ll}
    t^{(n)}, & \text{if $r$ and $r'$ are horizontal n.n.}\\
    t^{\prime (n)}, & \text{if $r$ and $r'$ are oblique n.n.}
  \end{array}
  \right.
\label{tani2p}
\end{eqnarray}
We consider the same fluxes and possible site-dependent potentials as in Sec.~\ref{subsec:2p}.  For example, at density $n=1/3$ we allow $\sqrt{3} \times \sqrt{3}$ pattern in the chemical potential.  At other densities, we can consider other appropriate CDW patterns.

One virtue of the 2-parton states is that they connect naturally to the decoupled chains limit.  Indeed, for $t^{\prime (n)} = 0$, ${\rm sign}[t^{(1)} t^{(2)}] = {\rm sign}[t^{(b)}] < 0$, the trial wavefunction Eq.~(\ref{det1det2}) on each chain reduces to
\begin{eqnarray}
\Psi_{\rm chain}(x_1, \dots, x_M) &\sim& e^{i\pi (x_1 + \dots + x_M)} 
\nonumber \\
&\times& 
\left\vert \prod_{i<j} \sin\frac{\pi (x_i - x_j)}{L} \right\vert^p
\label{Psichain}
\end{eqnarray}
with $p = p_1 + p_2$. The first factor gives correct Marshall sign for the 1D boson problem with hopping $t^{(b)} < 0$.  (To be more precise, we assume that the chain length $L$ is even and choose periodic or antiperiodic boundary conditions for the partons depending whether the number of bosons $M$ is odd or even.)  This is an accurate trial state in the full range of boson densities:  For $n=1/2$, with $p=2$ it reduces to the ground state of the Haldane-Shastry chain and is a good approximation to the ground state of the Heisenberg chain; for $n \to 0$, with $p=1$ it reproduces the nearly free fermion picture of the dilute gas of hard-core bosons; for varying $n$, by adjusting $p$ this state can capture varying Luttinger liquid exponents.  The 2-parton construction can thus provide a starting point for exploring what happens when the chains are coupled together.  As discussed in Sec.~\ref{subsec:2p}, the parton hopping between the chains with vector potentials satisfying Eq.~(\ref{t1t2}) can roughly capture the interchain boson hopping energy, while site-dependent chemical potentials can produce candidate CDW states.  We will present this in some detail for $n=1/3$ and $n=1/6$.

\textit{ED calculations}: To conclude the discussion of our anisotropic setups, we describe the supplementary ED calculations on the 36-site cluster with $J$ and $J'$ couplings. We compute a few lowest eigenvalues in each symmetry sector of the Hamiltonian with no twist and also eigenvalues in the zero momentum sector of the twisted Hamiltonian Eq.~(\ref{Htwisted}) with varying $\vec{Q}$.  At a given anisotropy and boson density, the minimum of these ED energies is taken to be the ground state energy.  Our ED calculations are restricted to $N_b \leq 12$.  The variational calculations are performed for the same 36-site cluster and also for larger systems.  

We now turn to the results of our anisotropic study.  For illustration, we present two boson densities.

\subsection{$n=1/3$}

\begin{figure}
  \centering
    \includegraphics[width=\columnwidth]{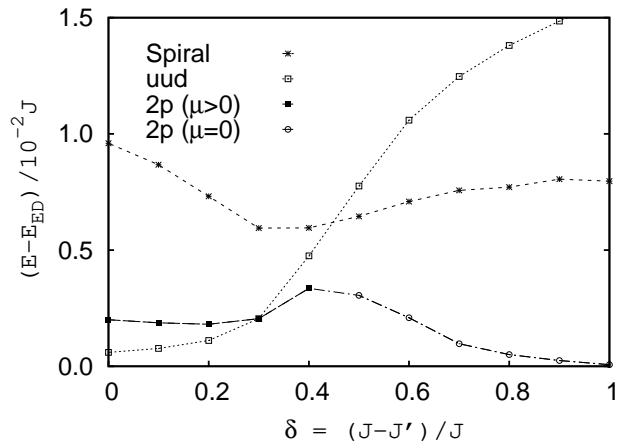}
  \caption{Comparison of spiral, permanent-type $uud$, and 2-parton trial energies (per bond) for 12 bosons on the anisotropic 36-site cluster.  The wavefunctions are anisotropic generalizations of those constructed in Sec.~\ref{sec:isotropic}.  For $\delta \lesssim 0.4$, the optimal 2-parton wavefunction has a non-zero chemical potential on sublattice $A$ and provides an alternative realization of the $uud$ state.  For $\delta > 0.4$, the chemical potential optimizes to zero, probably due to large finite-size gap for such anisotropy.  On a larger $24 \times 24$ cluster, the chemical potential remains non-zero up to $\delta \approx 0.7$, leading us to conjecture that in the thermodynamic limit the $uud$ phase persists all the way to $J'/J \to 0$.
}
  \label{fig:ani_energies_1_3}
\end{figure}

Figure~\ref{fig:ani_energies_1_3} shows the trial energies of $uud$, incommensurate spiral, and 2-parton wavefunctions at density $1/3$ on the 36-site cluster.  From the isotropic study, it is not surprising that the permanent-type $uud$ wavefunction remains a good candidate at low anisotropy.  As mentioned in Sec.~\ref{subsec:2p}, the 2-parton U1B wavefunction constructed with $0/\pi$ fluxes through triangles and a localizing chemical potential on one sublattice provides an alternative realization of the $uud$ state. For $\delta \gtrsim 0.3$, the 2-parton energy becomes lower than the permanent-type wavefunction but the $\sqrt{3} \times \sqrt{3}$ chemical potential (and therefore the $uud$ phase) persists up to $\delta \approx 0.4$.  Since our permanent wavefunction uses localized orbitals that only extend to nearest neighbour sites, it fails to capture longer range correlations in the chain direction expected in a more anisotropic system.  We would need to use more extended orbitals in the permanent, but we have not pursued this.  On the other hand, the 2-parton realization readily accommodates the lattice anisotropy via the parton hoppings, Eq.~(\ref{tani2p}), and provides a simple way to continue our study of the $uud$ state to larger anisotropy.

In the highly anisotropic region, we obtain good trial energies for the $6 \times 6$ system using the 2-parton wavefunction without the chemical potential.  However, if we consider the low energy cutoff due to the finite cluster size, it is clear that the study cannot resolve the true phase in the thermodynamic limit.  Specifically, in the decoupled chains limit, we obtain a very accurate wavefunction for two bosons on a 6-site chain by using antiperiodic boundary conditions for the partons, cf.~Eq.~(\ref{Psichain}).  The corresponding parton spectrum nicely accommodates two particles and has a large finite-size gap to next levels, which persists up to moderate interchain couplings.  While our 2-parton state by virtue of good fluxes naturally builds in good interchain exchange correlations, we cannot resolve the thermodynamic phase (e.g., the development of the $\sqrt{3} \times \sqrt{3}$ CDW) if the relevant energy scale is much lower than the finite-size gap.

To determine how far the $uud$ phase might extend into the anisotropic region, we repeat the 2-parton calculation on a large $24 \times 24$ cluster and find that the $\sqrt{3} \times \sqrt{3}$ chemical potential surprisingly remains non-zero up to $\delta \approx 0.7$. We also check that the parton spectrum for the optimal parameters is fully gapped and is connected to the strongly gapped CDW limit, so the trial wavefunction is indeed a valid charge-ordered Mott insulator of bosons as discussed in Sec.~\ref{subsec:2p}.  We thus conclude that the $uud$ state persists to rather strong anisotropy, albeit the CDW order becomes progressively weaker.  Interestingly, Ref.~\onlinecite{Starykh07} would predict the same $\sqrt{3} \times \sqrt{3}$ CDW order in the nearly decoupled chains limit at density $n=1/3$.  Combining with our variational work, this suggests that the $uud$ phase may in fact extend continuously from $\delta = 0$ up to $\delta = 1$ (See Fig.~\ref{fig:phase_diagram3}).  A rigorous confrontation to this conjecture could be provided, for example, by a systematic DMRG study of $3 \times L$ ladders at density $n=1/3$ varying $J'/J$ from $1$ to $0$ and monitoring the evolution of the $\sqrt{3} \times \sqrt{3}$ charge order.

While our study agrees with the observed plateau in $\textrm{Cs}_2\textrm{CuBr}_4$ ($\delta \approx 0.3$), it contradicts the absence of the plateau in $\textrm{Cs}_2\textrm{CuCl}_4$ ($\delta\approx 0.66$).  It is likely that residual interactions (e.g. such as Dzyaloshinskii-Moriya) have to be added to the Heisenberg model in order to describe the latter material,\cite{Starykh07} and they can change the energetics balance against the (very weak) $uud$ state in this highly anisotropic system.

\subsection{$n=1/6$}
\begin{figure}
  \centering
    \includegraphics[width=\columnwidth]{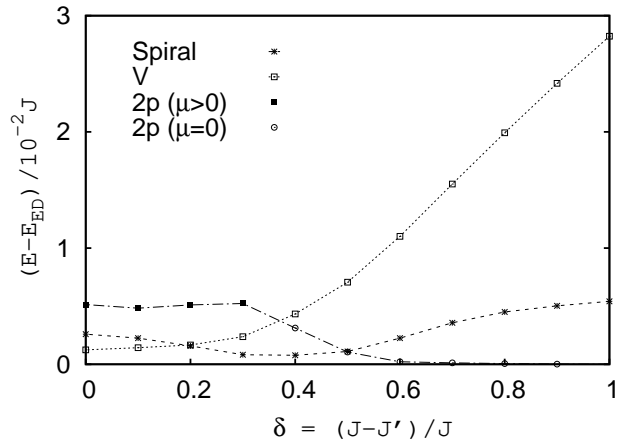}
  \caption{Comparison of spiral, commensurate $V_{\rm Bijl-Jastrow}$, and 2-parton trial energies as a function of anisotropy for 6 bosons on the $6 \times 6$ cluster.  The $E_{2p}$ values are obtained with higher chemical potential on one sublattice for $\delta \leq 0.3$ (in agreement with this state trying to capture 3-sublattice features near small anisotropy in this small sample).  We also performed a much larger study at $n=1/6$ comparing commensurate and incommensurate $V$ states, incommensurate spiral, and 2-parton states; from this study, the region of commensurate $V$ is actually quite small, while the incommensurate $V$ dominates over the spiral over a range $\delta \leq 0.3$ (see text for more details).}
  \label{fig:ani_energies_1_6}
\end{figure}

Figure~\ref{fig:ani_energies_1_6} shows the trial energies of commensurate $V_{\rm Bijl-Jastrow}$, incommensurate spiral [treated as described around Eq.~(\ref{Htwisted})], and 2-parton wavefunctions for 6 bosons on the $6 \times 6$ cluster.  At low anisotropy, the $V_{\rm Bijl-Jastrow}$ state is a good candidate.  For increasing anisotropy, a change to the incommensurate spiral is observed, which eventually loses to the ``quasi-1d'' phase represented by the 2-parton wavefunction with zero chemical potential.

In the highly anisotropic region, the figure shows remarkable agreement between the ED and the 2-parton energies, where we impose uniform $\pi/6$ and $5\pi/6$ flux per triangle for the $d_1$ and $d_2$ partons respectively (Chern-Simons state described in Sec.~\ref{subsec:2p}).  Despite the excellent agreement, we simply conclude that the highly anisotropic region is strongly dominated by quasi-1d physics and finite size effects.  Specifically, in the ED calculation on the 36-site cluster with 6 bosons, we find a non-degenerate ground state and a relatively large excitation energy gap.  We interpret this as follows.  In the decoupled chains limit, each chain contains one boson; for such a segment of length $L=6$, one expects a non-degenerate ground state with a large excitation gap due to finite size.  This gap persists as the chains are coupled, particularly because of some frustration present in the triangular lattice geometry.

We can similarly rationalize all our ED observations at other densities in the highly anisotropic limit. For example, for 7 bosons on the 36-site cluster, one of the chains now contains two bosons, and the ground state of the decoupled chains Hamiltonian is 6-fold degenerate due to 6 possible ways of choosing this chain.  The finite-size gaps ``protect'' this situation until the interchain coupling $J'$ becomes sufficiently large.  Such observations on the ED spectra show serious limitations of the small system study in the anisotropic model.  Going over all ED data for $N_b \leq 12$, we conclude that $\delta \gtrsim 0.5$ regime can be rationalized as such weakly-coupled finite chains, with no clear resolution of the ultimate state.  This is labeled as ``quasi-1d'' region in Fig.~\ref{fig:phase_diagram}.

One of the goals of the $6 \times 6$ study was to have ED reference for our trial states.  Having achieved some confidence in the good energetics of these states (despite their limitations), we now want to discuss variational results for larger sizes.  Specifically, on the 36-site cluster, we have not considered the possibility of incommensurate $V$ state: while we know how to accommodate the incommensurate spiral state, we do not have similar construction for the incommensurate coplanar state.  On the 36-site cluster, we see that incommensuration becomes important for $\delta \geq 0.2$.  In fact, as we discuss below, we think that for this density the $V$ state is probably incommensurate already for smaller anisotropy, but also extends to larger anisotropy in the competition against the spiral.

\subsection{Incommensurate $V$ versus spiral study at low to intermediate boson densities}
\label{subsec:VvsS}
In this section, we focus on the high-field regime where the incommensurate $V$ and spiral are the main competing candidates.  First, we briefly describe the relevant physical picture.
Beginning with the near-saturation limit, we consider a gas of free bosons hopping on the triangular lattice with the following
kinetic energy spectrum:
\begin{eqnarray}
  \epsilon_k = J\cos(k_x) + 2J'\cos\left(\frac{k_x}{2}\right) \cos\left(\frac{\sqrt{3}~k_y}{2}\right).
\end{eqnarray}
The band minima occur at $\vec{Q}=\pm\left(Q_x,0\right)$ with
\begin{equation}
  Q_x = 2 \ \mathrm{acos}(-J'/2J).
\label{Q}
\end{equation}
A condensation of bosons at these points gives rise to a degenerate manifold of states spanned by
\begin{eqnarray}
  \{(b_{\vec{Q}}^\dagger)^m (b_{-\vec{Q}}^\dagger)^{N_b-m} \vert 0 \ra ~;~ m = 0, 1, \hdots, N_b\}.
\end{eqnarray}
At low densities, the degeneracy is lifted by nearest neighbour repulsion. To see how this happens, we expand the interaction in terms of the two dominant spectral modes and then replace the operators by c-numbers\cite{Nikuni95,Veillette06}:
\begin{eqnarray}
  b_r &\sim& e^{i\vec{Q}\cdot\vec{r}}\ b_{\vec{Q}} + e^{-i\vec{Q}\cdot\vec{r}}\ b_{-\vec{Q}} ,\label{Eq:spectral_expansion}\\
  \hat{H}_{\rm int} &=& \sum_{\la rr'\ra} J_{rr'}\ b_r^\dagger b_r \ b_{r'}^\dagger b_{r'} \\
  &\sim& (J+2J') \left( \vert b_{\vec{Q}}\vert^2 + \vert b_{-\vec{Q}}\vert^2 \right)^2 
\nonumber\\
  && 
+\ 2\upsilon\ \vert b_{\vec{Q}}\vert^2\ \vert b_{-\vec{Q}}\vert^2 ,\\
  \upsilon &=& J\cos(2Q_x) + 2J'\cos(Q_x) .\label{Eq:nu}
\end{eqnarray}
The effect of nearest neighbour repulsion is determined by the sign of $\upsilon$.  For $J'/J < 0.39$, $\upsilon$ is positive and the incommensurate spiral (e.g., $\vert b_{\vec{Q}}\vert \neq 0$ and $\vert b_{-\vec{Q}}\vert = 0$) wins.  For $0.39 < J'/J \leq 1.59$, $\upsilon$ is negative and the incommensurate $V$ ordering ($\vert b_{\vec{Q}}\vert = \vert b_{-\vec{Q}}\vert$) becomes more stable.  The prediction from this approximate treatment is consistent with the spiral phase found in the nearly decoupled chains limit near saturation and in the highly anisotropic dilute boson study for the Cs$_2$CuCl$_4$;\cite{Starykh07,Veillette06} this is also consistent with the coplanar phase found in the isotropic dilute boson study.\cite{Nikuni95}  

In the above discussion, we have neglected the effect of hard core interaction.  Intuitively, this should be more important at higher density: The role of the hard core constraint is to prevent two bosons already in nearest neighbour contact from further occupying the same site, while at low density such contacts are avoided due to nearest neighbor repulsion.  To see whether the hard core interaction favors the $V$ or spiral phase, we expand the on-site repulsion energy in terms of the two spectral modes:
\begin{eqnarray}
  (b_r^\dagger b_r)^2 \sim \left( \vert b_{\vec{Q}}\vert^2 + \vert b_{-\vec{Q}}\vert^2 \right)^2 + \ 2\ \vert b_{\vec{Q}}\vert^2\ \vert b_{-\vec{Q}}\vert^2 . \label{eqn:hardcore}
\end{eqnarray}
From the positive sign in the second term, which dislikes the $V$,  we may expect the boundary between the $V$ and spiral phases to shift in favor of the spiral phase as the density increases.

\begin{figure}
  \centering
  \includegraphics[width=\columnwidth]{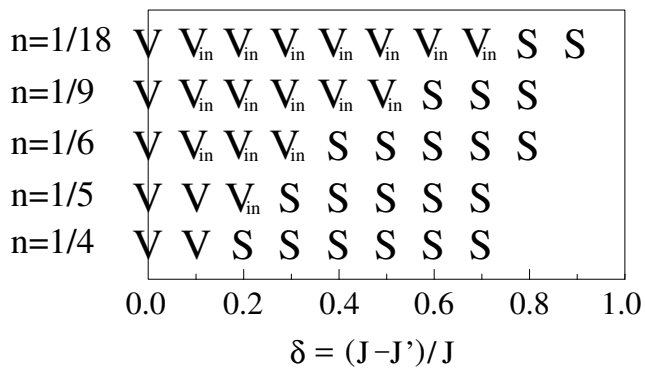}
  \caption{Variational phase boundary between $V$ and spiral (S) phases obtained from calculations on cluster sizes $18\times 18$, $51\times 20$, $27\times 20$, $44\times 20$, $47\times 20$, $38\times 10$, $39\times 10$, $53\times 10$, $60\times 10$, and $99\times 10$, where the corresponding anisotropy increases from 0 to 0.9 in equal intervals; the sizes are chosen so as to best accommodate the classical wavevector, Eq.~(\ref{Q}).  The $V$ phase remains commensurate at $\delta = 0.1$ for $n \geq 1/5$.   Note that at high anisotropy and particularly with increasing density, we find that the spiral loses to the 2-parton wavefunction, which we interpret as quasi-1D dominated regime.  
}
  \label{fig:VS_variational}
\end{figure}

To address the competition between these two phases quantitatively at finite density, we implement a variational study on $m \times n$ rectangular clusters such that a fitting wavevector $Q_x = 2\pi p/m$ is close to the spiral wavevector at each anisotropy (with appropriate integers $p$).  For the spiral phase, we use the same anisotropic wavefunction described earlier.  A candidate wavefunction for the incommensurate $V$ phase is constructed as follows:
\begin{eqnarray}
  |\psi_{V_{\rm in}} \ra &=& \left( e^{i\alpha} b_{\vec{Q}}^\dagger + e^{-i\alpha} b_{-\vec{Q}}^\dagger\right)^{N_b} \vert 0\ra \nonumber\\
  &=& \left( \sum_r\cos(\vec{Q}\cdot\vec{r}+\alpha)\ b_r^\dagger \right)^{N_b} \vert 0\ra.  
\end{eqnarray}
Note that for incommensurate $\vec{Q}$ the relative phase between $b_{\vec{Q}}^\dagger$ and  $b_{-\vec{Q}}^\dagger$ is not fixed, which we indicated with $\alpha$.  This is not important in an infinite system since $\vec{Q}\cdot\vec{r}$ visits all phases.  On the other hand, for commensurate $\vec{Q} = (4\pi/3, 0)$, $\alpha = 0$ and $\pi/2$ correspond to distinct $V$ and $\Psi$-type phases discussed in the isotropic case; both can be viewed as ``parent'' states for the incommensurate coplanar phase, but we will continue referring to the latter as $V$-type.

For ease of implementation, we use the same translationally invariant pseudopotentials given in Eq.~(\ref{Eq:ani_pseudopotential}).  The $V$ state has an incommensurate density wave and in principle allows more complicated pseudopotentials, so this choice probably biases slightly in favor of the spiral which has uniform boson density.  In all other respects, the physical setting and the variational freedom are very similar in our realizations of the spiral and $V$ states, and we think this study provides a fair comparison between the two phases even if the $Q_x$ may be slightly off and the Jastrow pseudopotentials are not the most general.

Fig.~\ref{fig:VS_variational} shows the result of our incommensurate $V$ versus spiral variational study.  The boundary between the two phases qualitatively agrees with our earlier argument, suggesting that the hard core repulsion is comparatively less important at low density.  We note that the obtained trial energies of the $V$ and spiral states are quite close (particularly at low density), hence the exact location of the phase boundary should not be taken as definitive.  Furthermore, the simple pseudopotential is clearly not optimal in the highly anisotropic regime, and eventually our spiral loses to the 2-parton states.  A more rigorous $V$ versus spiral variational study can be pursued by introducing more variational parameters into the Jastrow factor and employing systematic wavefunction optimization methods.\cite{Yunoki06}
The variational result that the coplanar phase extends to large anisotropy for fields near saturation is also in agreement with a recent dilute boson calculation extended to all $J'/J$.\cite{Starykh_unpub}

From the present results, we make an interesting observation that at the anisotropy relevant for \CsCuBr, $\delta \approx 0.3$, the transition occurs at density somewhere between $n=1/5$ (magnetization $0.6$ of saturation) and $n=1/6$ (magnetization $2/3$ of saturation).  The $V$ phase occupies the region near saturation, while the spiral occurs at lower magnetizations.  Thus, if the Heisenberg model is an adequate description, some of the features in the high field phase diagram of \CsCuBr\ may be due to the competing umbrella-type and coplanar states.\cite{Fortune09}

\subsection{Summary of anisotropic study}\label{sec:anisotropic_discussion}
\begin{figure}[t]
  \centering
  \includegraphics[width=\columnwidth]{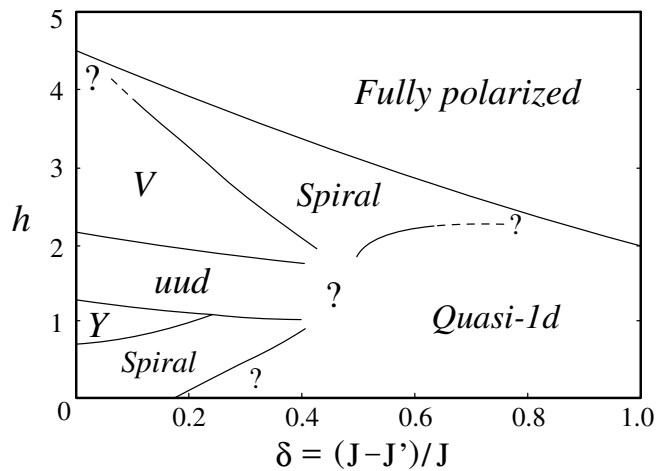}
  \caption{Variational phase diagram for the anisotropic $6 \times 6$ cluster in magnetic field.  In this study, we exclude incommensurate versions of $V$ and $Y$ states.  Due to finite size limitation, certain regions are marked as unresolved.  In addition, the region $\delta > 0.5$ is strongly dominated by quasi-1d physics, particularly for this small cluster study (see discussions in text).
}
  \label{fig:phase_diagram}
\end{figure}

Figure~\ref{fig:phase_diagram} summarizes a variational phase diagram obtained for the 36-site cluster considering all boson densities.  We label certain parts of the diagram with question marks or broken lines to indicate these regions as unresolved or less reliable.  The figure shows the $uud$ phase extending relatively far into the anisotropic region.  On both sides of the $uud$ phase, the commensurate coplanar phases remain stable over the incommensurate spiral for certain ranges of the anisotropy.  As the spatial anisotropy biases against the commensurate states, the actual $V$ and $Y$ regions are expected to be wider if the wavefunctions are generalized to incommensurate versions.  However, we exclude such extensions since they could not be accommodated on the $6 \times 6$ cluster.  For $\delta > 0.5$, our 2-parton trial energies are generally very good.  However, we think that this only indicates the onset of quasi-1d physics and strong finite-size effects as discussed earlier for the specific densities.

\begin{figure}[t]
  \centering
  \includegraphics[width=\columnwidth]{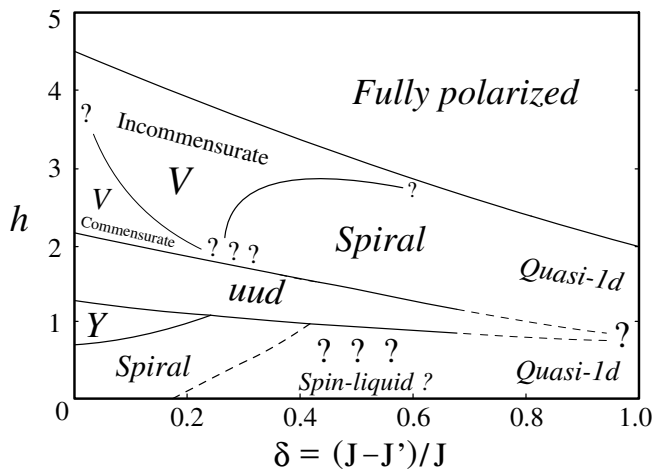}
  \caption{Schematic phase diagram for the anisotropic triangular lattice in magnetic field, combining $6\times 6$ study as well as larger cluster studies.  We suggest the possibility that the $uud$ plateau extends across the entire range of anisotropy.  The $V$ phase is commensurate (see Fig.~\ref{fig:spin_orderings}) near  the isotropic axis and the plateau but becomes incommensurate at moderate anisotropy and higher fields.  The highly anisotropic region is not well resolved in our variational study. 
The low field regime is also not studied thoroughly in this work, while previous studies\cite{Weng06,Yunoki06,Heidarian09a} suggest spin liquid state along $h=0$ and $\delta>0.2$.}
  \label{fig:phase_diagram3}
\end{figure}

Figure~\ref{fig:phase_diagram3} shows a schematic phase diagram based on the $6 \times 6$ anisotropic study as well as studies on larger clusters.  Here, we address a number of unresolved regions in Fig.~\ref{fig:phase_diagram}: limits of the $uud$ plateau, the boundary between $V$ and spiral, and the boundary between commensurate and incommensurate $V$.   
We find that the $uud$ phase extends much futher and may be even to all $\delta$. Also, a significant portion of the phase diagram at high fields is occupied by the incommensurate $V$ phase. 

We note that our work does not rule out other incommensurate phases found in the recent study by Alicea.\etal\cite{Alicea09}   For example, we were not able to come up with a good implementation of the incommensurate extension of the $Y$ state.
On the other hand, we did try Bijl-Jastrow-type wavefunctions for distorted umbrella states discussed in Ref.~\onlinecite{Alicea09}, which are commensurate supersolids with incommensurate spiral phase angles.  On the $36$-site cluster, these trial states optimized to the incommensurate spiral with uniform boson density, but we have not explored this thoroughly on larger clusters.  Overall, our results are more conclusive at low densities and much less at high densities between 1/3 and 1/2.

\section{Summary and discussion}\label{sec:conclusion}

We studied the Heisenberg antiferromagnet on the spatially anisotropic triangular lattice in the field from a variational perspective.  On the isotropic lattice, we constructed a very simple and physically transparent permanent-type wavefunction for the $uud$ state at density $1/3$.  This is a Mott insulator of bosons where we accurately included small charge fluctuations by using appropriate localized boson orbitals.  The remarkable trial energy suggests that such approach may be useful in other Mott insulator contexts.  Next, we obtained natural extensions to nearby $V$ and $Y$ supersolid phases respectively for $n\lesssim 1/3$ and $n\gtrsim 1/3$, where the physics remains strongly influence by the proximity to $n=1/3$.  By connecting to a Bijl-Jastrow-type candidate wavefunction at low density, the coplanar $V$ phase extends to all $n<1/3$ (i.e., up to the saturation field in the spin model language).  Note, however, that at very low density another coplanar state ($\Psi$-type) is expected to be very close,\cite{Nikuni95} and we cannot resolve between the two.
On the higher density side of the plateau (i.e., at lower fields), the permanent-type $Y$ wavefunction performs well near the plateau but narrowly loses to the Huse-Elser spiral candidate at densities close to half-filling (zero field).  The latter result is consistent with other recent works.\cite{Wang09, Jiang09, Heidarian09b}
 
The success of our isotropic study encouraged us to extend it to the anisotropic lattice.  At density $n=1/3$, we begin with the permanent-type realization of the $uud$ and then connect to a conceptually similar but technically different 2-parton realization at higher anisotropy. Surprisingly, we found that the $uud$ phase remains stable over a large range of anisotropy.  In conjunction with the same CDW phase found in the decoupled chains limit,\cite{Starykh07} we suggest that the $uud$ phase may in fact extend across the entire range of anisotropy.  This conjecture can be tested more rigorously using a DMRG study on finite-width strips.

In the low boson density region (i.e., at high fields), the Bijl-Jastrow-type $V$ commensurate supersolid wavefunction is smoothly connected to the incommensurate version.  This state competes with the incommensurate spiral, and we can accurately compare the two.  We found that the incommensurate $V$ state has lower energy in a large region of the phase diagram, extending up to a fairly large value of anisotropy in the very dilute regime (i.e., close to the saturation fields).  On the other hand, the $V$ phase remains commensurate near the isotropic axis and the plateau.

In the high density regime (i.e., at low fields), we attempted to construct an incommensurate $Y$ candidate using a Bijl-Jastrow-type wavefunction but found that this construction performs poorly.  This low field region at moderate to high lattice anisotropy calls for more comprehensive investigation.

One of the goals we had was to explore possible new plateaus in the high field regime of $\textrm{Cs}_2\textrm{CuBr}_4$.  We have learned that the phase diagram is already very rich even without considering any additional plateaus.  Nevertheless, for several densities such as $1/6$, $2/9$, and $1/4$ we implemented permanent-type wavefunctions for various proposed CDW from Ref.~\onlinecite{Fortune09} as well as for some additional stripe-like orderings, and inevitably found that either $V$ or spiral has lower energy.  Our earlier $uud$ study showed that the 2-parton construction can also be useful for studying CDW phases; however, similar implementations at the above densities again failed to reveal any stable charge ordering.  This suggests that any such order, if present at all, is likely to be very weak. 

One of the findings from our study is that for the \CsCuBr\ anisotropy, the system is in the incommensurate coplanar phase close to the saturation fields,\cite{Starykh_unpub} and there may be a transition to the non-coplanar spiral state at lower fields; this could be responsible for one of the features in the $\textrm{Cs}_2\textrm{CuBr}_4$ experiment.  We cannot exclude other more complex cascade of phases.  Furthermore, additional residual interactions not treated here may be important for understanding the phases of $\textrm{Cs}_2\textrm{CuBr}_4$ in the field.  This remains a fascinating open problem.

\appendix

\section{Motivation for $V_{\rm perm}$ wavefunction for $n \lesssim 1/3$, Eq.~(\ref{eqn:Vperm_matrix})}
\label{app:Vperm}
We begin with the $uud$ state with $N/3$ bosons, Eq.~(\ref{eqn:uud_wf}), and put $N_h = N/3 - N_b$ holes in a ``hole orbital'' $\phi_h(R)$,
\begin{eqnarray}
|\Psi\ra &=& \left(\sum_R \phi_h(R) b_R \right)^{N_h}\ |\psi_{uud}\ra 
~.
\end{eqnarray}
For a boson configuration
\begin{eqnarray}
|\eta\ra &=& b_{r_1}^\dagger \dots b_{r_{N_b}}^\dagger |0\ra ~,
\end{eqnarray}
we obtain an amplitude,
\begin{widetext}
\begin{equation}
\la\eta | \Psi\ra = 
\sideset{}{^\prime}\sum_{R_1 \dots R_{N_h}} \phi_h(R_1) \dots \phi_h(R_{N_h})
\ \mathrm{Perm}
\begin{pmatrix}
\phi_1^\mathrm{loc}(r_1) & \dots & \phi_{N/3}^\mathrm{loc}(r_1) \\
\vdots & \ddots & \vdots \\
\phi_1^\mathrm{loc}(r_{N_b}) & \dots & \phi_{N/3}^\mathrm{loc}(r_{N_b}) \\
\phi_1^\mathrm{loc}(R_1) & \dots & \phi_{N/3}^\mathrm{loc}(R_1) \\
\vdots & \ddots & \vdots \\
\phi_1^\mathrm{loc}(R_{N_h}) & \dots & \phi_{N/3}^\mathrm{loc}(R_{N_h})
\end{pmatrix}
\approx
\mathrm{Perm}
\begin{pmatrix}
\phi_1^\mathrm{loc}(r_1) & \dots & \phi_{N/3}^\mathrm{loc}(r_1) \\
\vdots & \ddots & \vdots \\
\phi_1^\mathrm{loc}(r_{N_b}) & \dots & \phi_{N/3}^\mathrm{loc}(r_{N_b}) \\
c_1 & \dots & c_{N/3} \\
\vdots & \ddots & \vdots \\
c_1 & \dots & c_{N/3}
\end{pmatrix}
\end{equation}
\end{widetext}
The primed sum indicates that $R_1, \dots, R_{N_h}$ need to be different from each other and from all $r_1, \dots, r_{N_b}$.  Close to the $1/3$ plateau, the density of holes is small, and we can approximately replace the restricted sum by an unrestricted sum.  Performing independent summations over $R_1, \dots, R_{N_h}$ gives the last expression in the form of a single permanent, where
\begin{equation}
c_j = \sum_R \phi_h(R) \phi_j^\mathrm{loc}(R) 
\end{equation}
is an ``overlap'' of the $\phi_h$ and $\phi_j^\mathrm{loc}$ orbitals.  On physics grounds, the hole orbital $\phi_h$ needs to respect the symmetries of the $uud$ state ($\phi_h = {\rm const}$ over the $A$ sublattice).  In this case, $c_j$ is independent of $j$, and up to a normalization constant we can replace all matrix elements in the last $N_h$ rows by $1$.  The approximate single permanent form is a valid variational wavefunction by itself, and this is the state we use in the main text and call $V_{\rm perm}$.

\section{Correlation functions of permanent-type states}\label{app:corr}
\begin{figure*}
  \centering
  \begin{tabular}{ccc}
    \includegraphics[trim=0 10 0 0]{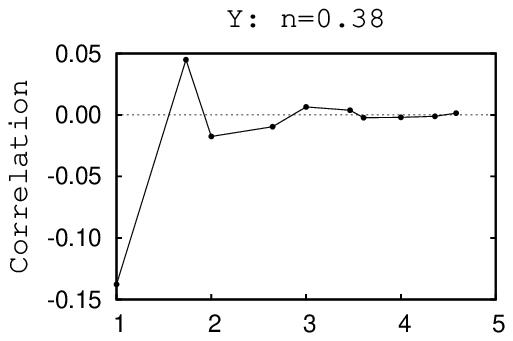} &
    \includegraphics[trim=10 10 10 0]{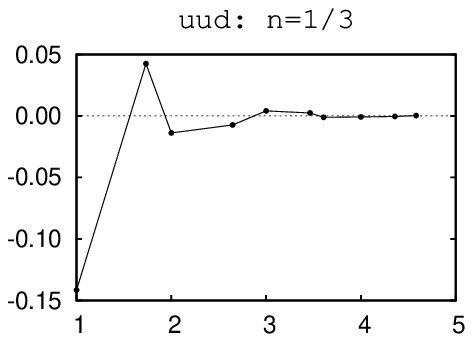} &
    \includegraphics[trim=0 10 0 0]{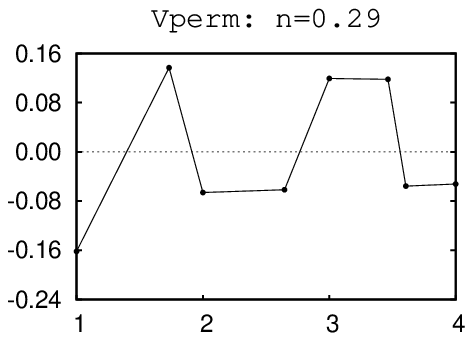} \\
    \includegraphics[trim=0 0 0 5]{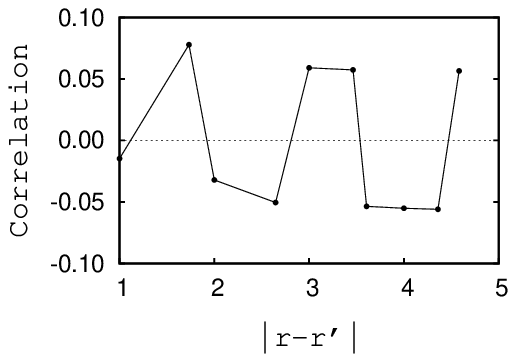} &
    \includegraphics[trim=10 0 10 5]{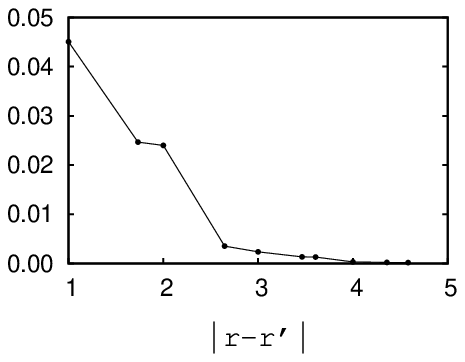} &
    \includegraphics[trim=0 0 0 5]{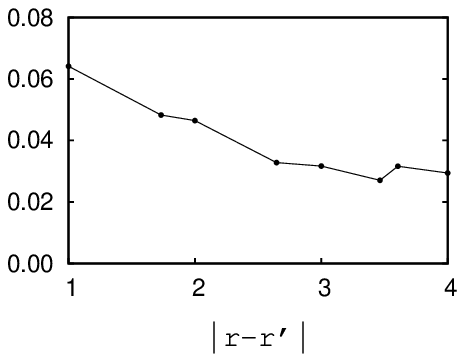}
  \end{tabular}
  \caption{Boson correlation $\la b_r^\dagger b_{r'}\ra$ as a function of real space distance $\vert r - r'\vert$.  Top row: $r$ belongs to sublattice $A$ (where $n_A > n_B = n_C$) while $r'$ includes sites on all three sublattices.  Bottom row: $r$ belongs to sublattice $B$ while $r'$ includes sites on sublattices $B$ and $C$. Hexagonal clusters with 84, 144, and 48 sites are respectively used for $Y$ (left column), $uud$ (middle), and $V_{\rm perm}$ (right) calculations.}
  \label{fig:correlation}
\end{figure*}

We calculate the order parameters and correlation functions in the permanent-type trial states, since we do not have much experience with such wavefunctions for Mott insulators or supersolids.  The non-permanent $V$ and spiral trial states are more obvious constructions, and therefore omitted.

For $Y$, $uud$, and $V_{\rm perm}$ trial states, a three sublattice modulation is observed in the CDW order parameter $\la n_r \ra$.  As expected, the density structure factor $\la n_{-q}n_q\ra$ reveals sharp peaks near the reciprocal vectors $\vec{Q} = \pm(\frac{4\pi}{3}, 0)$ for all three trial states.

Figure~\ref{fig:correlation} shows the correlation functions $\la b_r^\dagger b_{r'}\ra$ between two sites for these trial states. The $uud$ state has rapidly decaying correlations between any two sites, which is expected in this Mott insulator state.  For the $Y$ supersolid state, the correlations decay rapidly when at least one site lies on the higher density sublattice $A$ (i.e., as if this sublattice is Mott-insulating), while they are long-ranged when both sites reside on the $BC$ honeycomb sublattice.  The signs of the correlations are positive for all pairs of $B$-$B$ or $C$-$C$ sites, and negative for all $B$-$C$ pairs, which is consistent with the $Y$ spin order shown in Fig.~\ref{fig:spin_orderings}. Finally, for the $V_{\rm perm}$ trial state, long-ranged correlation exists between any two sites on the lattice. The signs are negative for all pairs of $A$-$B$ or $A$-$C$ sites, and positive for all $B$-$C$ pairs (as well as $A$-$A$, $B$-$B$, and $C$-$C$ pairs), which is consistent with the $V$ spin order shown in Fig.~\ref{fig:spin_orderings}.  Thus, we have verified our intuition about the physical properties of these states.\\

\acknowledgments

We would like to thank J. Alicea and O. Starykh for many stimulating discussions and for reading and commenting on the manuscript.  This research is supported by the A. P. Sloan Foundation and the National Science Foundation through grant DMR-0907145.


\end{document}